\documentclass[review]{elsarticle}

\usepackage{lineno,hyperref}
\usepackage{amsmath,graphicx}
\usepackage{mathtools}
\usepackage{cancel}
\usepackage{comment}
\usepackage{capt-of}
\usepackage{color}

\modulolinenumbers[500]

\journal{Nuclear Physics A}

\begin{document}

\begin{frontmatter}

\title{Energy loss of heavy quarks and $B$ and $D$ meson spectra in PbPb collisions at 
LHC energies}

\author[ad1]{Kapil Saraswat}
\author[ad2,ad3]{Prashant Shukla\corref{ca}}
\cortext[ca]{Corresponding author}
\ead{pshuklabarc@gmail.com}
\author[ad2]{Vineet Kumar}
\author[ad1]{Venktesh Singh}

\address[ad1]{Department of Physics, Banaras Hindu University, Varanasi 221005, India.}
\address[ad2]{Nuclear Physics Division, Bhabha Atomic Research Centre, Mumbai 400085, India.}
\address[ad3]{Homi Bhabha National Institute, Anushakti Nagar, Mumbai 400094, India.}

\begin{abstract}

  We study the production and evolution of charm and bottom quarks in hot partonic
medium produced in heavy ion collisions. The heavy quarks loose energy in the medium
which is reflected in the transverse momentum spectra of heavy mesons.
 The collisional energy loss of heavy quarks has been calculated using QCD calculations.  
The radiative energy loss is obtained using two models namely reaction operator formalism 
and generalized dead cone approach. 
  The nuclear modification factors, $R_{AA}$ as a function of transverse momentum by including 
shadowing and energy loss are calculated for $D^{0}$ and $B^{+}$ mesons in PbPb collisions 
at $\sqrt{s_{NN}}$ = 5.02 TeV and for $D^{0}$ mesons at $\sqrt{s_{NN}}$ = 2.76 TeV and
are compared with the recent measurements.
  The radiative energy loss from generalized dead cone approach alone is sufficient 
to produce measured $D^{0}$ meson $R_{AA}$ at both the LHC energies.
  The radiative energy loss from reaction operator formalism plus collisional 
energy loss gives good description of $D^{0}$ meson $R_{AA}$.
   For the case of $B^{+}$ meson, the radiative energy loss from generalized dead cone 
approach plus collisional energy loss gives good description of the CMS data. 
The radiative process is dominant for charm quarks while for the bottom, both
the radiative process and the elastic collisions are important.
  
\end{abstract}


\end{frontmatter}

\linenumbers

\section{Introduction}

  The heavy ion collisions at Relativistic Heavy Ion Collider (RHIC) and Large 
Hadron Collider (LHC) are performed to create and characterize Quark Gluon Plasma (QGP). 
 The properties of QGP are studied through variety of probes accessible in these 
experiments \cite{quarkmatter:2014}. 
 The heavy (charm and bottom) quarks are the best probes of the transport properties of the 
medium. Since the heavy quarks are produced in hard partonic interactions in heavy ion collisions,
their initial momentum distribution can be calculated from pQCD 
\cite{Beenakker:1988bq, Beenakker:1990maa, Kumar:2012qx}.
While traversing the hot/dense medium formed in the collisions, these quarks loose energy 
due to the elastic collisions with the plasma constituents and/or by radiating a gluon. 
 There are several formulations to calculate collisional 
\cite{Bjorken:1982tu, Braaten:1991jj, Braaten:1991we, Peshier:2006hi, Peigne:2008nd} as well 
as radiative energy loss \cite{Armesto:2003jh, Armesto:2004vz, Gyulassy:2000fs, Djordjevic:2003zk}.
For a review of many of these formalism see Ref.~\cite{Jamil:2010te, Prino:2016cni}. 
The collisional energy loss dominates at low parton energy but the radiative energy loss 
dominates over the collisional energy loss at high parton energy \cite{GayDucati:2005jq}.
  A recent work in Ref.~\cite{Kochelev:2015jba} finds significant non-perturbative contribution to 
collision energy loss accompanying a pion production in quark - gluon - pion interaction.

 The ALICE experiment measured nuclear modification factor ($R_{AA}$) \cite{ALICE:2012ab} 
and elliptic flow \cite{Abelev:2013lca,  Abelev:2014ipa} of $D$ mesons in 
PbPb collisions at $\sqrt{s_{NN}}$ = 2.76 TeV.
 Many transport models employing heavy quark dynamics have been used to interpret this data
\cite{Uphoff:2012gb, Beraudo:2014boa,  Cao:2014dja, Bratkovskaya:2015foa} which we summarize in 
the following.
   A Boltzmann Approach to MultiParton Scatterings (BAMPS) \cite{Uphoff:2012gb} is a transport 
model which describes the $D$ meson data very well. The model lacks the radiative energy loss which is 
accounted for by multiplying the collision cross-section by 3.5.
  The POWLANG is a Monte Carlo model \cite{Beraudo:2014boa} where the initial heavy quarks pairs are
produced by POWHEG-BOX and their propagation in hydrodynamically expanding medium is simulated 
through Langevin equation.
 The hydrodynamic model from Ref.~\cite{Cao:2014dja} uses a modified Langevin equation with 
terms for collisional and radiative interactions. The transport coefficients
are then tuned to produce the $D$ meson $R_{AA}$ at $\sqrt{s_{NN}}$ = 2.76 TeV.
  In Parton-Hadron-String Dynamics (PHSD) transport approach \cite{Bratkovskaya:2015foa}, 
the initial charm quarks are produced by tuned PYTHIA which scatter with the off shell partons 
whose masses and widths are given by the Dynamical Quasi Particle Model (DQPM). 
In this model, radiative process is suppressed due to large gluon mass in DQPM.
  HYDJET++ model \cite{Lokhtin:2008xi,Lokhtin:2016xnl} is a Monte Carlo model which
includes collision energy loss calculated in the high momentum limit and 
the radiative energy loss is obtained by generalization of 
BDMPS (Baier, Dokshitzer, Mueller, Peigne and Schiff) model based on dead cone approximation.

  The $R_{AA}$ of $B$ meson via its decay to $J/\psi$ was measured by the CMS experiment 
\cite{CMS:2012vxa}. The measurements of both $D$ and $B$ at LHC and $D$ at RHIC are used 
to constrain energy loss formalisms in our simple hydrodynamic model  
by modifying the transverse momentum ($p_T$) spectra of heavy quarks due to
collision and radiative energy loss \cite{Saraswat:2015ena}.  
  ALICE and CMS recently updated $D$ meson $R_{AA}$ in extended $p_T$ \cite{Adam:2015nna} and 
centrality \cite{Adam:2015jda} range in PbPb collisions at $\sqrt{s_{NN}}$ = 2.76 TeV.
CMS has published good quality measurements of $R_{AA}$ of $D^{0}$ \cite{CMS:2016nrh} and
$B^{+}$ \cite{CMS:2016jya} mesons and elliptic flow, $v_2$ of $D^{0}$ mesons \cite{CMS:2016jtu}
in PbPb collision at $\sqrt{s_{NN}}$ = 5.02 TeV. 
 These new LHC data can be used to test various models of heavy quark energy loss.

 In this work, first we calculate the $p_T$ spectra of heavy mesons in pp collision
at $\sqrt{s}$ = 5.02 TeV
using pQCD model \cite{Beenakker:1988bq, Beenakker:1990maa} and make a comparison with 
$D^{0}$ and $B^{+}$ meson measurements of CMS. The radiative energy loss of charm and bottom quarks 
are calculated using reaction operator formalism DGLV (Djordjevic, Gyulassy, Levai and Vitev) 
\cite{Gyulassy:2000fs, Djordjevic:2003zk, Wicks:2005gt} and generalized dead cone approach 
\cite{Saraswat:2015ena, Abir:2012pu}. The collisional energy loss is calculated using
Peigne and Peshier formalism \cite{Peigne:2008nd}. The nuclear modification 
factors including shadowing and energy loss are calculated for $D^{0}$ and $B^{+}$ mesons 
in PbPb collision at $\sqrt{s_{NN}}$ = 5.02 TeV and are compared with the CMS measurements.
We also calculate $R_{AA}$ for $D^{0}$ meson in PbPb collision at $\sqrt{s_{NN}}$ = 2.76 TeV
to compare with the updated data from ALICE and CMS.

\section{Heavy Quark Production}

The heavy quarks are produced by the processes 
$q + \bar{q} \rightarrow Q + \bar{Q}$ and $g + g \rightarrow Q + \bar{Q}$
in the pp collisions as  
\begin{equation}
p(P_{1}) +  p(P_{2}) \rightarrow Q(p_{1}) + \bar{Q}(p_{2}) + X~.
\label{heavyflavourproductionone}
\end{equation}
The hadronic kinematic variables are 
\begin{eqnarray}
S &=& (P_{1} + P_{2})^{2} ,~  \nonumber \\ 
T_{1} &=& (P_{1} - p_{1})^{2} - m^{2}  = - \sqrt{S}~m_{T}~e^{y}~, \nonumber \\ 
U_{1} &=& (P_{1} - p_{2})^{2} - m^{2}  = - \sqrt{S}~m_{T}~e^{-y}~, 
\end{eqnarray}
where $y$  is the rapidity, $m_{T}(=\sqrt{p^{2}_{T} + m^{2}})$ is the transverse mass,
$p_{T}$ is the transverse momentum and $m$ is the mass of heavy quark.
The cross section for the process given in Eq.~\ref{heavyflavourproductionone} is 
\begin{eqnarray}
S^{2} \frac{d^{2}\sigma(S, T_{1}, U_{1})}{dT_{1} dU_{1}} = k \sum_{i, j} \int \frac{dx_{1}}{x_{1}}
\int\frac{dx_{2}}{x_{2}} f^{p}_{i}(x_{1}, Q^{2}) f^{p}_{j}(x_{2}, Q^{2}) 
s^{2} \frac{d^{2}\sigma_{i j} (s, t_{1}, u_{1})}{dt_{1} du_{1}}~.
\label{hadronicxsectionone}
\end{eqnarray}
Here, $s = x_{1} x_{2} S,~ t_{1} = x_{1} T_{1},~ u_{1} = x_{2} U_{1}$ are partonic variables.
The functions $f^{p}_{i}(x_{1}, Q^{2})$ denote the parton distribution functions (PDFs) in nucleons.
We take $Q = m_T$ and the $k$ factor is adjusted to reproduce the data. The mass of charm (bottom) 
quark is taken as 1.50 (5.0) GeV.
The Born cross section in 4 dimensions for $g g$ and $q \bar{q}$ 
interaction can be written in the following form 
\begin{eqnarray}
s^{2} \frac{d^{2}\sigma_{ij}}{dt_{1}~du_{1}} = \delta(s + t_{1} + u_{1}) \times 
\sigma_{ij}(s, t_{1}, u_{1})~.
\label{borncrosssectionhadron}
\end{eqnarray}
From Eqs.~\ref{hadronicxsectionone} and \ref{borncrosssectionhadron}
\begin{eqnarray}
\frac{d^{2}\sigma_{pp}}{dp^{2}_{T} dy} =
\frac{k}{S}  \sum_{i, j} \int^{1}_{x_{1-}} \frac{dx_{1}}{x_{1}} \Bigg(-\frac{1}{t_{1}}\Bigg)~ 
f^{p}_{i}(x_{1}, Q^{2}) ~  f^{p}_{j}(x_{2}, Q^{2})~ \sigma_{ij}(s, t_{1}, u_{1}). 
\label{borncrosssectionhadron1}
\end{eqnarray}
Here, $x_{1_{-}} = - U_{1}/(S + T_{1}) ~{\rm{and}}~ x_{2} = - x_{1} T_{1}/(x_{1} S + U_{1})$~.
The Born cross sections $\sigma_{ij}$ calculated upto LO are given in the appendix.

CT10 parton density functions \cite{Lai:2010vv} are used in the present calculations. 
The spatially dependent EPS09s sets \cite{Helenius:2012wd} are used to calculate the 
modifications of the PDFs inside nucleus. The differential cross section including 
nuclear shadowing effect corresponding to a centrality class between impact 
parameters $b_1$ and $b_2$ is calculated as 
\begin{eqnarray}
\frac{d^{2}\sigma_{\rm{sh}}(b_{1}, b_{2})}{dp^{2}_{T}~dy} &=&  
  \frac{k}{S}~  \sum_{i, j}  \int^{1}_{x_{1_{-}}} \frac{dx_{1}}{x_{1}}
\Big(-\frac{1}{t_{1}}\Big) ~ \sigma_{i j }(s, t_{1}, u_{1})~
\frac{1}{A B} \sum^{4}_{n, m =0} T^{n m}_{A B} (b_{1}, b_{2}) \nonumber \\
&~&  c^{i}_{n}(x_{1}, Q^{2})~f^{A}_{i}(x_{1}, Q^{2}) ~ 
 c^{j}_{m}(x_{2}, Q^{2})~f^{B}_{j}(x_{2}, Q^{2})~,  
\label{pbpbyield_with_centrality}
\end{eqnarray}
where the bound state PDFs $f^{A, B}_{i,j}$, the function $T^{n m}_{A B}$ and 
the coefficients $c^{i,j}_{n,m}$ are given in EPS09s sets \cite{Helenius:2012wd}. 
The spectrum in PbPb collisions is then obtained by including the momentum
loss $\Delta p_{T}$ in the $p_{T}$ spectrum given in Eq.~\ref{pbpbyield_with_centrality}.


  Single heavy meson production cross sections for both the pp and PbPb collisions are
obtained by convoluting the heavy quark production cross section with the fragmentation 
function $D^{h}_{Q}(z)$ \cite{Arleo:2004gn} as
\begin{equation}
\frac{d^{2}\sigma^{h}}{d(p^{h}_{T})^{2}dy} = f_{\rm{meson}} ~\int^{1}_{0} dz ~\frac{D^{h}_{Q}(z)}{z^{2}}~
\frac{d^{2}\sigma}{dp^{2}_{T}dy}.
\end{equation}
Here, $z = p^{h}_{T}/p_{T}$ and $f_{\rm{meson}}$ is the fragmentation fraction for the heavy meson. 
We take $f_{\rm{meson}}$ as 0.557 for $D^{0}$ meson \cite{Abelev:2012vra, Nakamura:2010zzi} 
and 0.402 for $B^{+}$ meson \cite{CMS:2016jya}. Peterson fragmentation function is used for 
$D^{h}_{Q}(z)$ \cite{Peterson:1982ak} which is given as follows  
\begin{equation}
D^{h}_{Q}(z) = \frac{N}{z\Big[1-\frac{1}{z} - \frac{\epsilon_{Q}}{(1-z)}\Big]^{2}}.
\end{equation}
We take $\epsilon_{c}$ = 0.016 and $\epsilon_{b}$ = 0.0012 and $N$ is the normalization constant. 

Finally, the nuclear modification factor $R_{AA}$ is calculated as
\begin{eqnarray}
R_{AA} (p^{h}_{T}, b_{1}, b_{2}) = \frac{d^{2}\sigma^{h}_{PbPb}(p^{h}_{T}, b_{1}, b_{2})}{d(p^{h}_{T})^{2} dy} 
\Bigg/\int^{b_{2}}_{b_{1}}~  d^{2}b ~T_{AA}~ \frac{d^{2}\sigma^{h}_{pp}(p^{h}_{T})}{d(p^{h}_{T})^{2} dy}~.
\end{eqnarray}
Here, $T_{AA}$ is the nuclear overlapping function.

\section{\bf Heavy Quark Energy Loss}
\label{HeavyQuarkEnergyLoss}
 
 For the collisional energy loss we use the formalism of 
Peigne and Peshier (PP) \cite{Peigne:2008nd}.
The radiative energy loss is calculated using the reaction operator formalism 
(DGLV) ~\cite{Gyulassy:2000fs, Djordjevic:2003zk, Wicks:2005gt} and using the generalized 
dead cone approach \cite{Abir:2012pu}.
 The DGLV formalism is based on a systematic expansion of the energy loss in terms of the 
number of the scatterings experienced by the propagating parton.
  In the single hard scattering limit, only the leading term in the expansion is included.
The Generalised dead cone approach is an extension of the Gunion Bertsch formalism \cite{Gunion:1981qs}.
  The Gunion Bertsch formula for light quarks energy loss was extended to
heavy quarks by introducing the mass in the matrix element but only within the small angle
approximation \cite{Dokshitzer:2001zm}. Due to this mass effect, the soft gluon emission
from a heavy quark was suppressed in comparision to that from a light quark which is
known as the dead cone effect.
  In the generalized dead-cone approach the probability of gluon emission off a heavy quark 
is obtained by relaxing some of the constraints such as the gluon emission angle and
the scaled mass of the heavy quark with its energy.
Using the same assumptions as generalized dead cone approach \cite{Abir:2012pu}
we calculated the energy loss expression \cite{Saraswat:2015ena} given as 
\begin{equation}
\frac{dE}{dx}=24~\alpha^{3}_{s}~\rho_{QGP}~\frac{1}{\mu_{g}}~\Big(1-\beta_{1}\Big)
~\Bigg(\sqrt{\frac{1}{(1-\beta_{1})}~\log\Big(\frac{1}{\beta_{1}}\Big)}-1 \Bigg)
~\mathcal F(\delta)~~.
\end{equation}
Here,
\begin{eqnarray}
\mathcal F(\delta) &=& 2\delta-\frac{1}{2}~\log\Bigg(
\frac{1+\frac{M^2}{s}~e^{2\delta}}{1+\frac{M^2}{s}~e^{-2\delta}}\Bigg)-
\Bigg(\frac{\frac{M^2}{s}~\sinh(2\delta)}
{1+2~\frac{M^2}{s}\cosh(2\delta)+\frac{M^4}{s^{2}}}\Bigg)~~, \nonumber \\
\delta &=& \frac{1}{2}~\log\Bigg[\frac{1}{(1-\beta_{1})}~\log\Big(\frac{1}
{\beta_{1}}\Big)~\Bigg(1+\sqrt{1-\frac{(1-\beta_{1})}{\log(\frac{1}{\beta_{1}})}} 
\Bigg)^{2} \Bigg]~, \nonumber \\
s &=& 2E^2+2E\sqrt{E^2-M^2}-M^2~, ~~~\beta_{1} = \mu^{2}_{g}/(C~E~T),  \nonumber \\
C &=& \frac{3}{2}-\frac{M^{2}}{4~E~T}+\frac{M^{4}}{48~E^{2}~T^{2}~\beta_{0}}~
\log\Big[\frac{M^{2}+6~E~T~(1+\beta_{0})}{M^{2}+6~E~T~(1-\beta_{0})}\Big],  \nonumber \\
\beta_{0} &=& \sqrt{1-\frac{M^{2}}{E^{2}}}~~ ,~~ 
\rho_{QGP} = \rho_{q}+\frac{9}{4}~\rho_{g} ~, \nonumber \\
\rho_{q} &=& 16 T^{3} \frac{1.202}{\pi^{2}} ~~, ~ \rho_{g} = 9 N_{f} T^{3} \frac{1.202}{\pi^{2}}~.
\end{eqnarray}
$\mu_{g}=\sqrt{4\pi \alpha_{s} T^{2}\Big(1+N_{f}/6\Big)}$ is the Debye screening mass, 
$T$ is the temperature of the QGP medium, $\alpha_{s}(=0.3)$ is the fine structure 
splitting constant for strong interaction and $N_f(=3)$ is the number of quark flavours.

\section{\bf Model For QGP Evolution}

  The average distance $L$ travelled by the heavy quark in the plasma is obtained as per 
the method described in Ref.~\cite{Saraswat:2015ena}. If  the velocity of the heavy quark 
is $v_{T}=p_{T}/m_{T}$, the effective path length $L_{eff}$ is obtained as
\begin{equation}
L_{eff} = {\rm min} \Big[L ,~ v_{T}~ \times ~\tau_{f} \Big].
\end{equation}
 The temperature as a function of proper time is obtained for  
each centrality bin in an isentropic cylindrical expansion scenario with 
the Lattice QCD and hadronic resonance equations of states \cite{Kumar:2014kfa}. 
We calculate the energy loss as a function of proper time which is then averaged 
over the temperature evolution.
 The measured values of $dN/d\eta$ at $\sqrt{s_{NN}}$ = 5.02 TeV \cite{Adam:2015ptt} 
and at $\sqrt{s_{NN}}$ = 2.76 TeV \cite{Aamodt:2010cz} are used as inputs for a given 
centrality to calculate the initial temperature. The initial and freezs-out times
are taken as 0.3 and 6 fm/$c$ respectively. 
  Various parameters used in our model for different centralities 
such as average value of impact parameter $<b>$, maximum value $b_{\rm{max}}$, number of 
participants $N_{\rm{part}}$ and the measured $dN/d\eta$ are given in the 
Table~\ref{qgp_evolution_model_parameter} along with the calculated values of $L$ and initial 
temperature $(T_{0})$.

\begin{table}[ht]
\caption{Parameters of QGP evolution model }
\begin{center}
\begin{tabular}{| c | c | c | c | c | c | c | c |} 
\hline
$\sqrt{s_{NN}}$ & Centrality & $<b>$ & $b_{\rm{max}}$ & $N_{\rm{part}}$ &$\frac{dN}{d\eta}$ & $L$    & $T_{0}$  \\ 
   (TeV)       & class ($\%$)     & (fm)  &    (fm)      &              &                   & (fm)   & (GeV) \\ \hline\hline 
   5.02        & 0-10     & 3.34  &     5.0      &  359         & 1749              &  5.74  & 0.508 \\ \hline 
   5.02        & 0-100    & 9.65  &     22.0     &  114         & 436               &  4.18  & 0.469 \\ \hline 
   2.76        & 0-10     & 3.44  &     5.0      &  356         & 1449              &  5.73  & 0.467 \\ \hline 
   2.76        & 0-100    & 9.68  &     22.0     &  113         & 363               &  4.16  & 0.436 \\ \hline 
\end{tabular}
\end{center}
\label{qgp_evolution_model_parameter}
\end{table}

\section{Results and Discussions}

\begin{figure}[htp]
\centering
\includegraphics[width=0.60\linewidth]{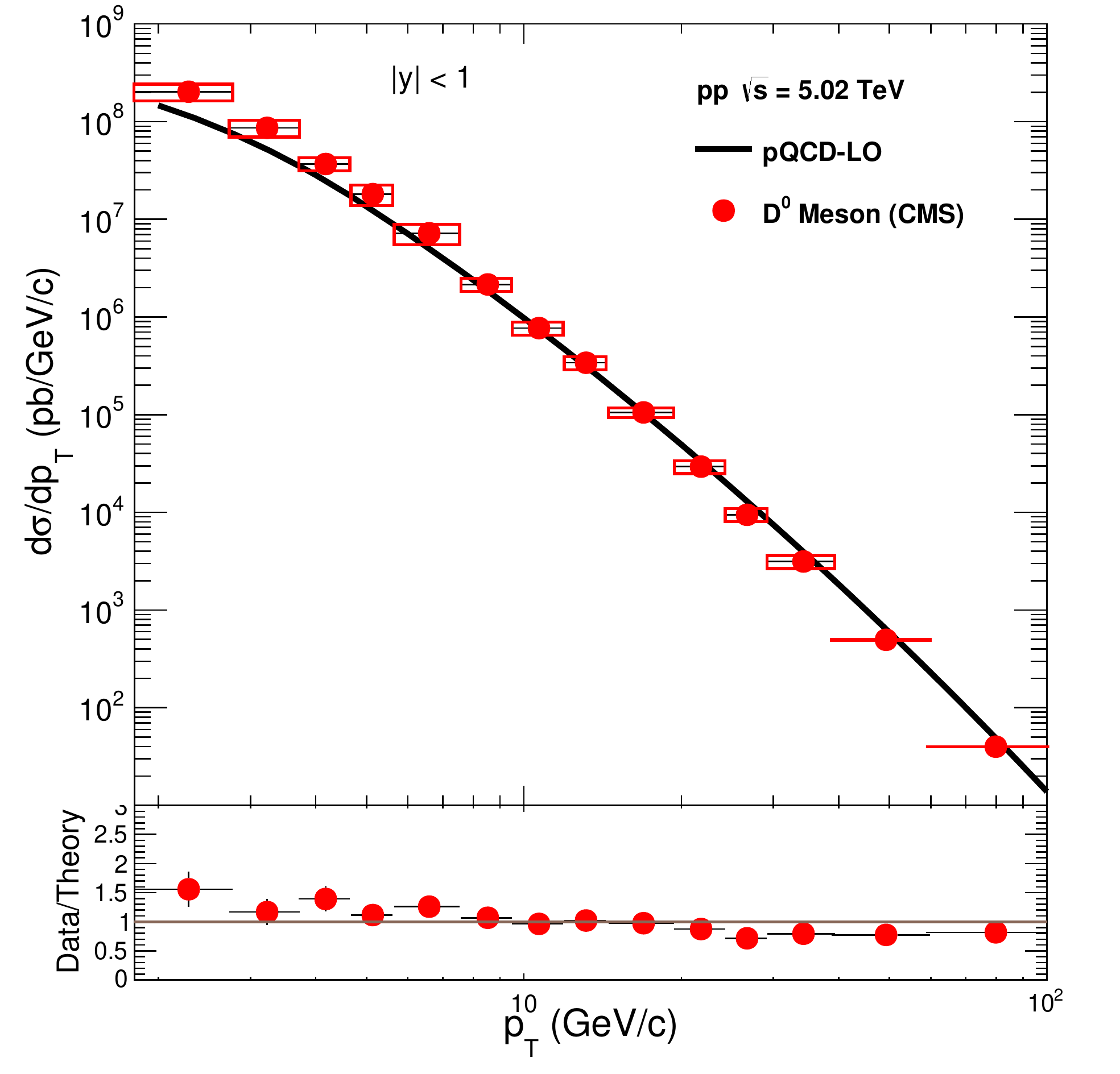}
\caption{(color online): The pQCD LO calculation of differential cross section 
of $D^{0}$ mesons as a function of the transverse momentum $p_{T}$, in pp collision 
at $\sqrt{s}$= 5.02 TeV.
The calculations are compared with the CMS data of $D^{0}$ mesons \cite{CMS:2016nrh}.}
\label{figure1_dmeson_yield}
\end{figure}

Figure \ref{figure1_dmeson_yield} shows the pQCD LO calculation of differential 
cross section of $D^{0}$ mesons as a function of transverse momentum $p_{T}$, in 
pp collision at $\sqrt{s}$= 5.02 TeV compared with the CMS measurements \cite{CMS:2016nrh}. 
The calculation with factor $k$ = 4 gives good description of the data.

\begin{figure}[htp]
\centering
\includegraphics[width=0.60\linewidth]{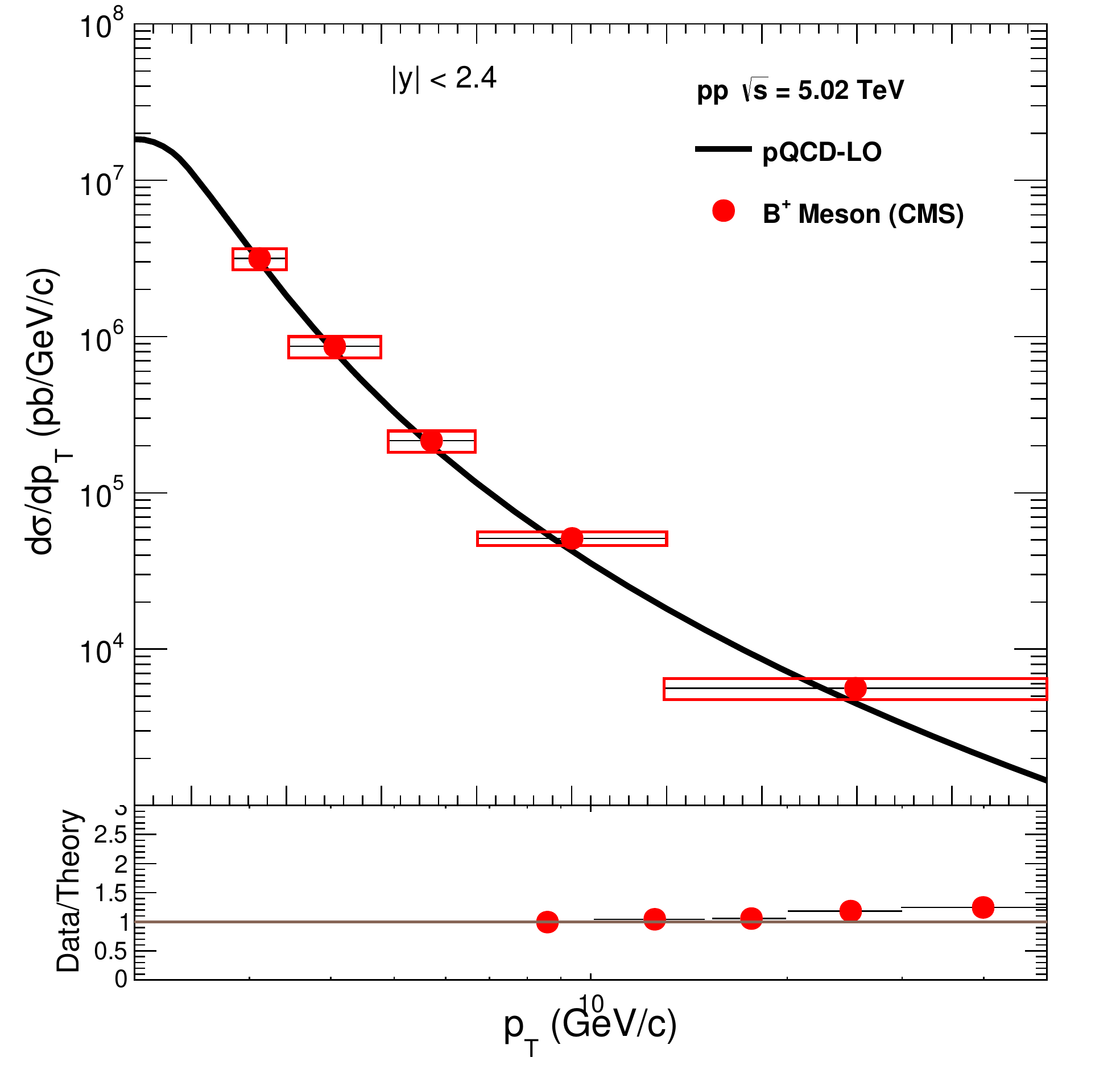}
\caption{(color online): The pQCD LO calculation of differential cross section 
of $B^{+}$ mesons as a function of the transverse momentum $p_{T}$, in pp collision 
at $\sqrt{s}$= 5.02 TeV.
The calculations are compared with the CMS data of $B^{+}$ mesons \cite{CMS:2016jya}.}
\label{figure2_bplusmeson_yield}
\end{figure}

Figure \ref{figure2_bplusmeson_yield} shows the pQCD LO calculation of differential 
cross section of $B^{+}$ mesons as a function of the transverse momentum $p_{T}$, in 
pp collision at $\sqrt{s}$= 5.02 TeV compared with the CMS measurements \cite{CMS:2016jya}. 
The calculation with factor $k$  = 5  gives good description of the data.

\begin{figure}[htp]
\centering
\includegraphics[width=0.60\linewidth]{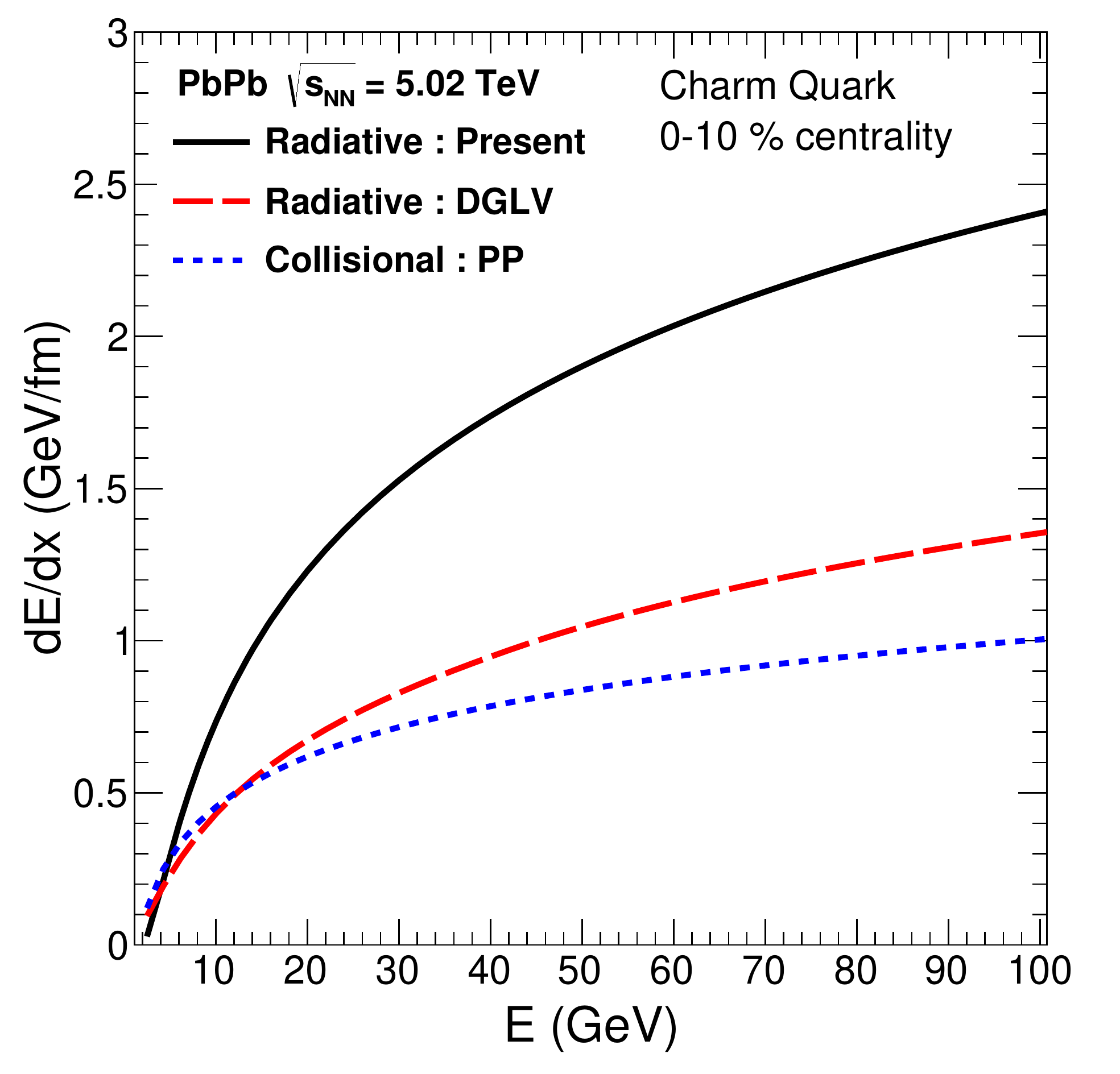}
\caption{(color online): The energy loss $dE/dx$ as a function of energy $E$ of charm quark 
obtained using PP, DGLV and Present calculation in 0 - 10 $\%$ centrality region for PbPb 
collision at $\sqrt{s_{NN}}$ = 5.02 TeV.}
\label{figure3_energyloss_0_10}
\end{figure}

\begin{figure}[htp]
\centering
\includegraphics[width=0.60\linewidth]{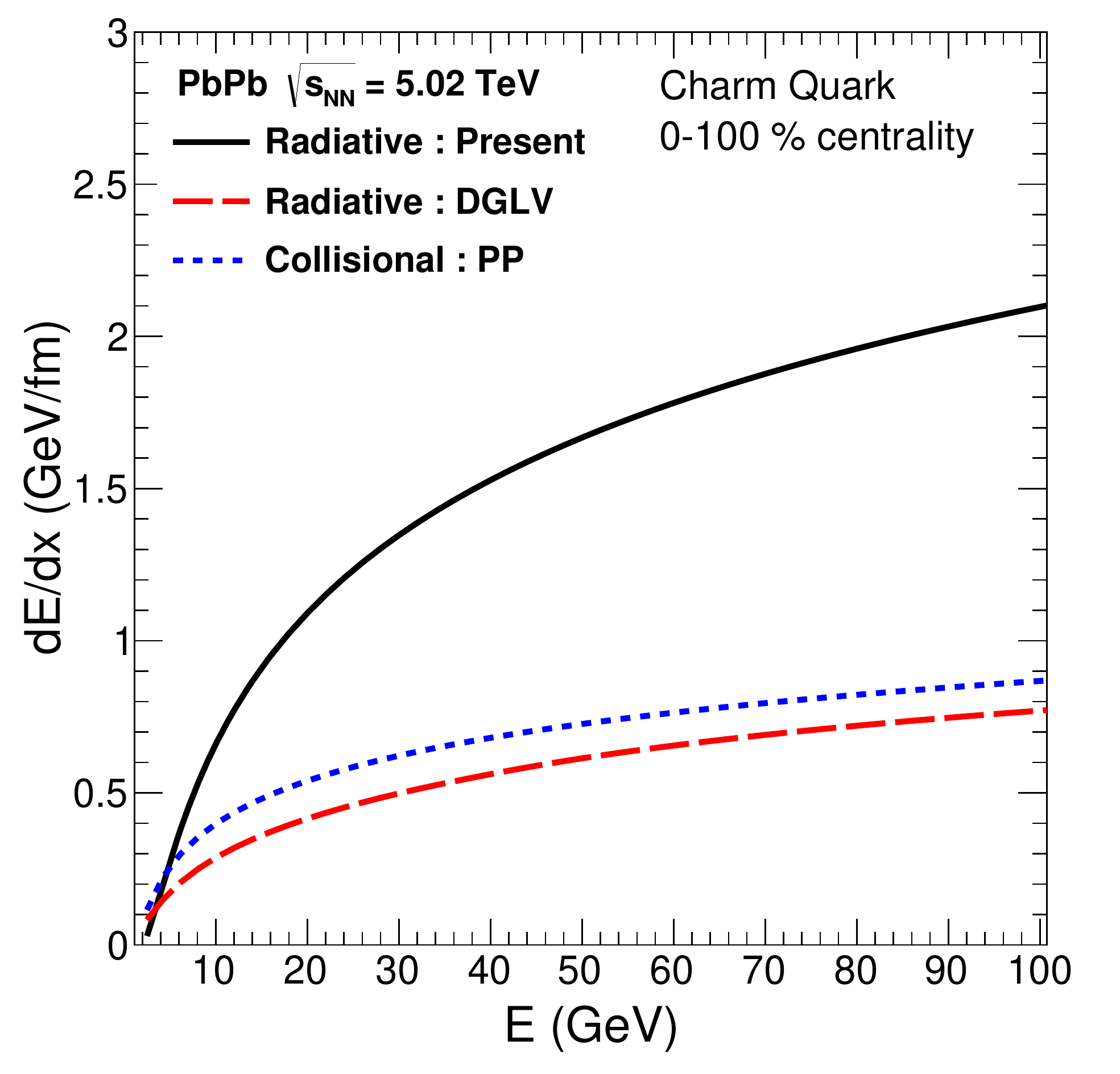}
\caption{(color online): The energy loss $dE/dx$ as a function of energy $E$ of charm quark 
obtained using PP, DGLV and Present calculation in 0 - 100 $\%$ centrality region for PbPb 
collision at $\sqrt{s_{NN}}$ = 5.02 TeV.}
\label{figure4_energyloss_0_100}
\end{figure}

Figure \ref{figure3_energyloss_0_10} shows the energy loss of charm quark as a function of 
quark energy for the case of 0 - 10 $\%$ central PbPb collision at $\sqrt{s_{NN}}$ = 5.02 TeV 
calculated using PP, DGLV and Present formalisms. The radiative energy loss calculated by
present approach 
is larger than that by DGLV. The collisional energy loss calculated by PP formalism is 
less than the radiative energy loss calculation. 
  Figure \ref{figure4_energyloss_0_100} is the same as Fig.~\ref{figure3_energyloss_0_10} 
but for the case of minimum bias PbPb collisions. 

\begin{figure}[htp]
\centering
\includegraphics[width=0.60\linewidth]{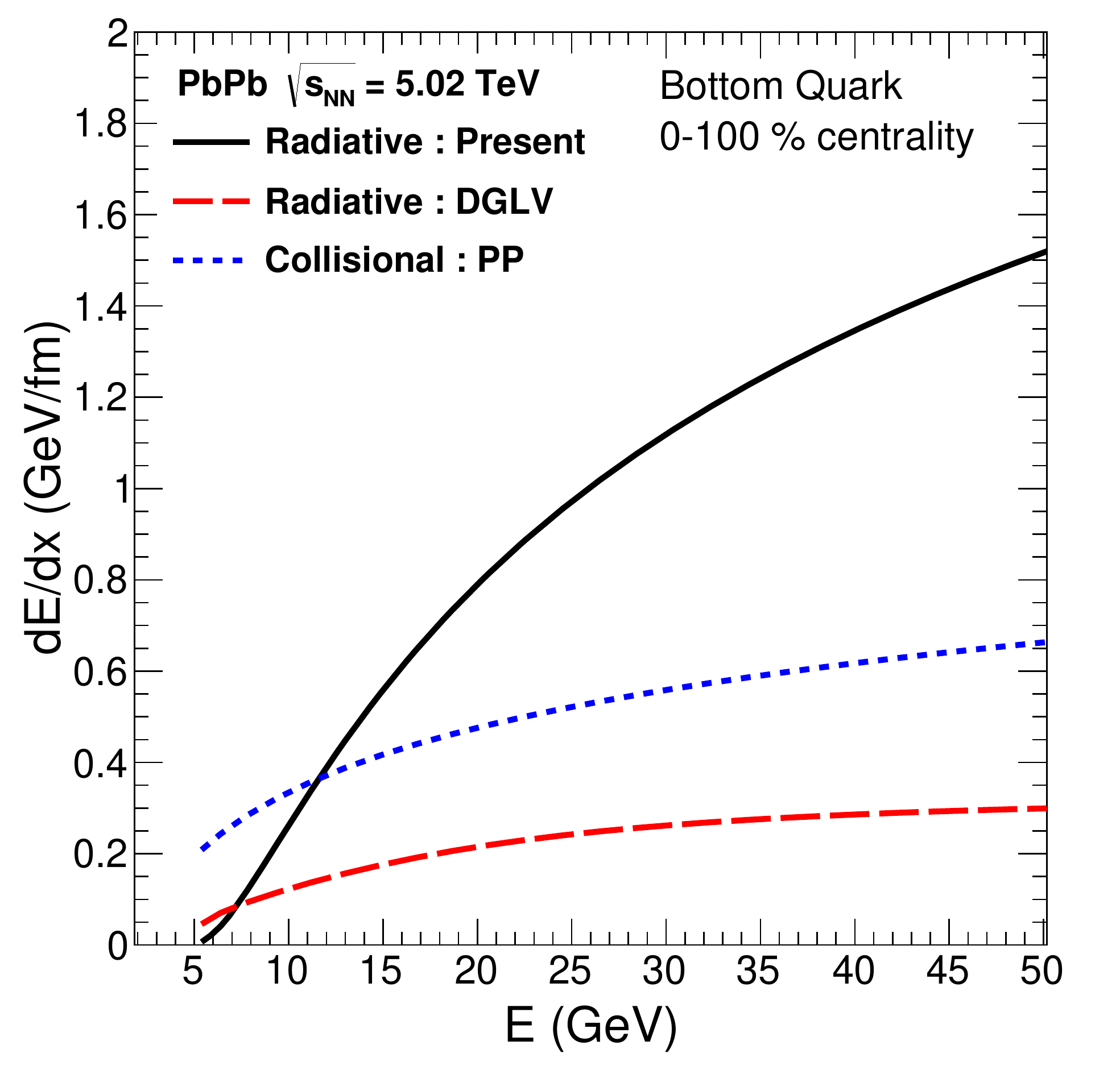}
\caption{(color online): The energy loss $dE/dx$ as a function of energy $E$ of bottom quark 
obtained using PP, DGLV and Present calculation in 0 - 100 $\%$ centrality region for PbPb 
collision at $\sqrt{s_{NN}}$ = 5.02 TeV.}
\label{figure5_bottom_energyloss_0_100}
\end{figure}

Figure \ref{figure5_bottom_energyloss_0_100} shows the energy loss of bottom quark as a function 
of quark energy for the minimum bias PbPb collisions at $\sqrt{s_{NN}}$ = 5.02 TeV using PP, DGLV 
and Present formalisms. The radiative energy loss calculated by present approach is larger than 
that by DGLV. The collisional energy loss for the bottom quarks is significant as compared to 
the radiative energy loss.

 The radiative energy loss calculated by the generalized dead cone approach is larger
  than the energy loss calculated by DGLV. This arises due to different kinematic cuts
  used in the two formalisms. Namely, in the DGLV formalism the gluon emission 
is constrained only to the forward angles $\theta~ <~ \pi/2$, where as in the
generalized dead cone approach, full range of $\theta$ is taken care of.

\begin{figure}[htp]
\centering
\includegraphics[width=0.60\linewidth]{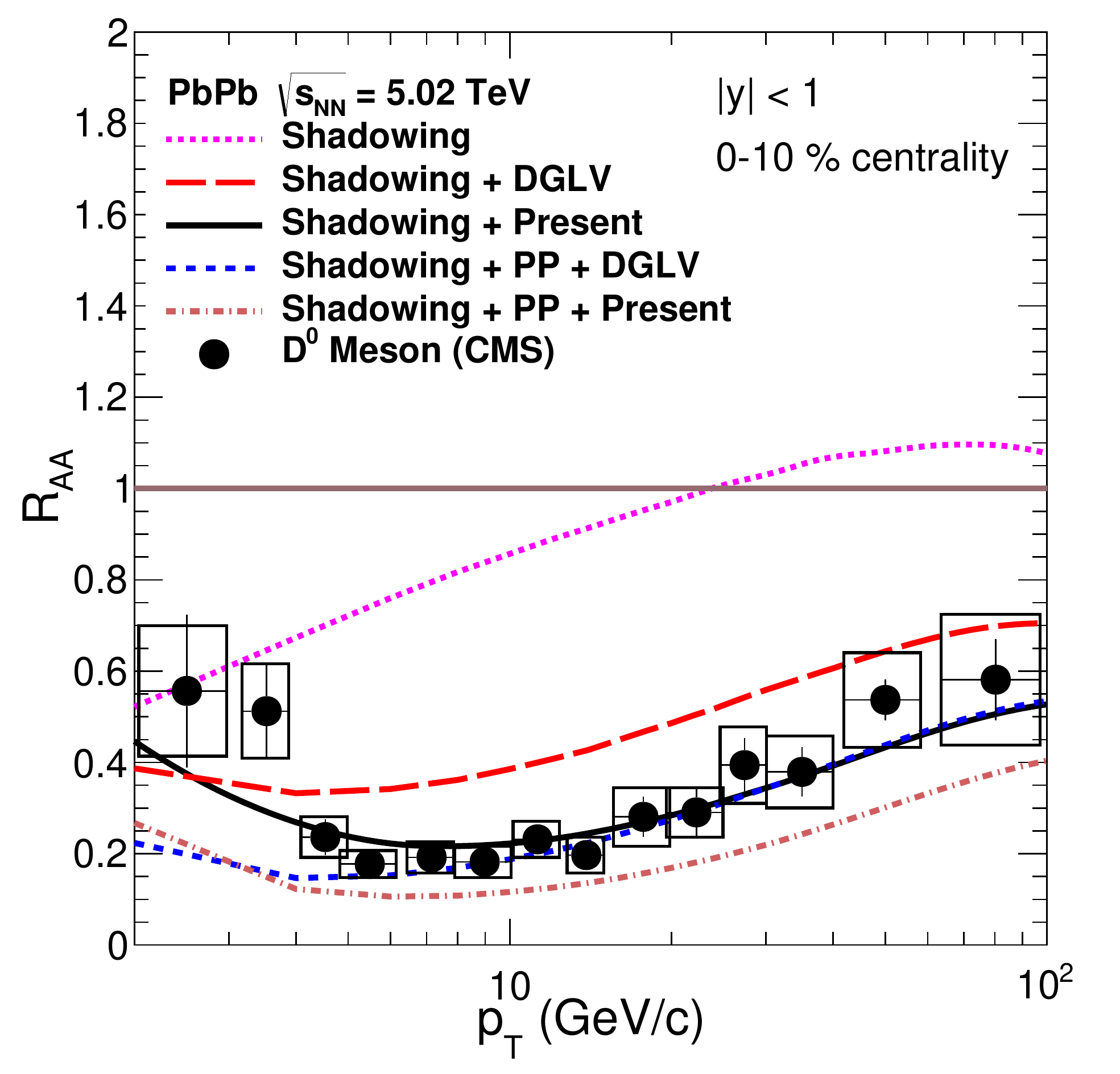}
\caption{(color online): Nuclear modification factor $R_{AA}$ of $D^{0}$ 
meson as a function of the transverse momentum $p_{T}$, obtained using energy loss 
(DGLV, Present, PP + DGLV and PP + Present calculation) and shadowing in PbPb collision at 
$\sqrt{s_{NN}}$= 5.02 TeV. The calculations are compared with the CMS data\cite{CMS:2016nrh}.}
\label{figure6_raa_0_10}
\end{figure}

Figure \ref{figure6_raa_0_10} shows the nuclear modification factor $R_{AA}$ of 
$D^{0}$ as a function of the transverse momentum $p_{T}$, obtained 
by including shadowing and energy loss (DGLV, Present, PP + DGLV and PP + Present calculations) 
for 0 - 10 $\%$ central PbPb collision at $\sqrt{s_{NN}}$= 5.02 TeV.
The calculations are compared with the CMS data \cite{CMS:2016nrh}.
We observe that the radiative 
energy loss by present formalism reproduces the data without adding collisional energy loss.
 The radiative energy loss by DGLV added to the collisional energy loss by 
PP describes the CMS data at high $p_{T}$ range. The radiative energy loss by present formalism 
addded to the collisional energy loss by PP formalism overestimates the measured suppression of 
$D^{0}$ meson.

\begin{figure}[htp]
\centering
\includegraphics[width=0.60\linewidth]{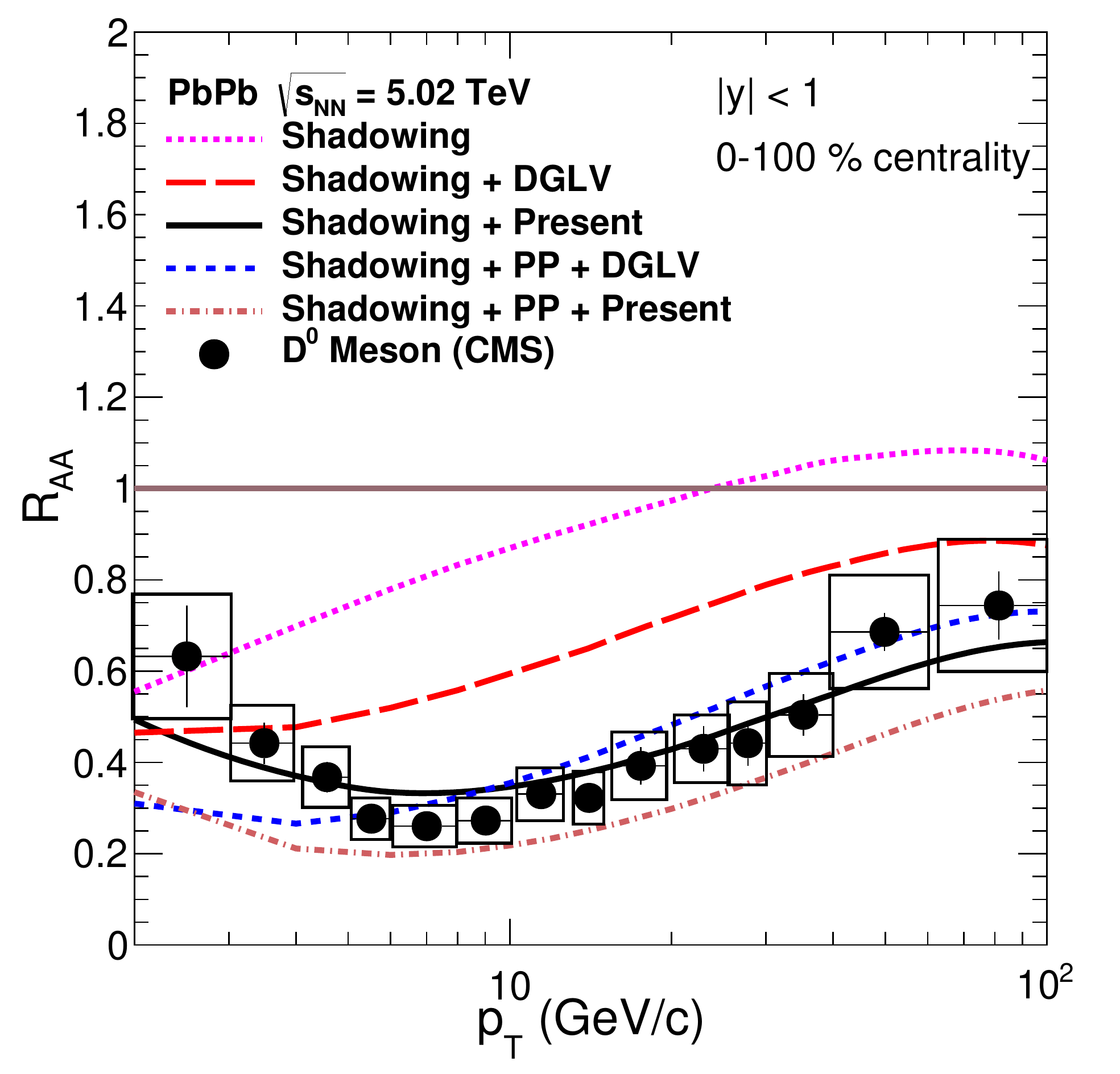}
\caption{(color online): Nuclear modification factor $R_{AA}$ of $D^{0}$ 
meson as a function of the transverse momentum $p_{T}$, obtained using energy loss 
(DGLV, Present, PP + DGLV and PP + Present calculation) and shadowing in PbPb collision 
at $\sqrt{s_{NN}}$= 5.02 TeV.
The calculations are compared with the CMS data \cite{CMS:2016nrh}.}
\label{figure7_raa_0_100}
\end{figure}

Figure \ref{figure7_raa_0_100} shows the nuclear modification factor $R_{AA}$ of 
$D^{0}$ as a function of the transverse momentum $p_{T}$, obtained by including shadowing 
and energy loss (DGLV, Present, PP + DGLV, PP + Present calculations) for the minimum bias PbPb 
collision at $\sqrt{s_{NN}}$= 5.02 TeV.
The calculations are compared with the CMS data \cite{CMS:2016nrh}. The radiative energy
loss by present formalism describes the CMS 
data within the uncertainties of the data. The sum of radiative and collisional energy 
loss (PP + DGLV) gives good description of the data at high $p_T$. The radiative energy 
loss by present formalism addded to the collisional energy loss by PP formalism overestimates 
the measured suppression of $D^{0}$ meson.

\begin{figure}[htp]
\centering
\includegraphics[width=0.60\linewidth]{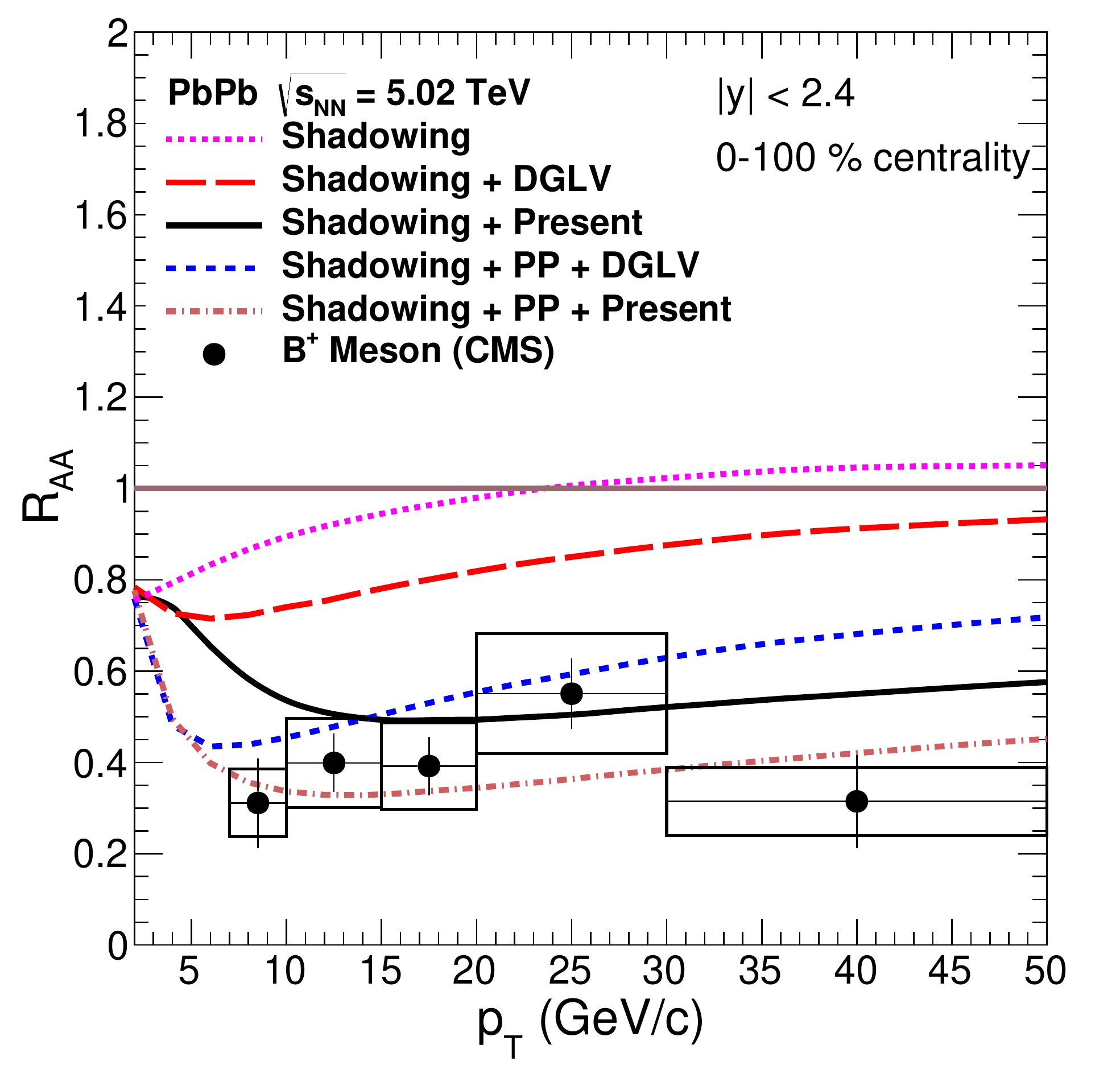}
\caption{(color online): Nuclear modification factor $R_{AA}$ of $B^{+}$ meson as a 
function of the transverse momentum $p_{T}$, obtained using energy loss 
(DGLV, Present, PP + DGLV and PP + Present calculation) and shadowing in PbPb 
collision at $\sqrt{s_{NN}}$= 5.02 TeV.
The calculations are compared with the CMS data of $B^{+}$ mesons \cite{CMS:2016jya}.}
\label{figure8_bmeson_raa_0_100}
\end{figure}

Figure \ref{figure8_bmeson_raa_0_100} shows the nuclear modification factor $R_{AA}$ of 
$B^{+}$ as a function of the transverse momentum $p_{T}$, obtained by including shadowing 
and energy loss (DGLV, Present, PP + DGLV and PP + Present calculations) for the minimum 
bias PbPb collision at $\sqrt{s_{NN}}$= 5.02 TeV.
The calculations are compared with the CMS data \cite{CMS:2016nrh}.
The sum of the radiative energy loss by present formalism and collisional 
energy loss by PP formalism describes the CMS data within the uncertainties of the data. The sum 
of the radiative energy loss by DGLV formalism and collisional energy loss by PP formalism 
underestimates the $B^{+}$ meson suppression.

\begin{figure}[htp]
\centering
\includegraphics[width=0.60\linewidth]{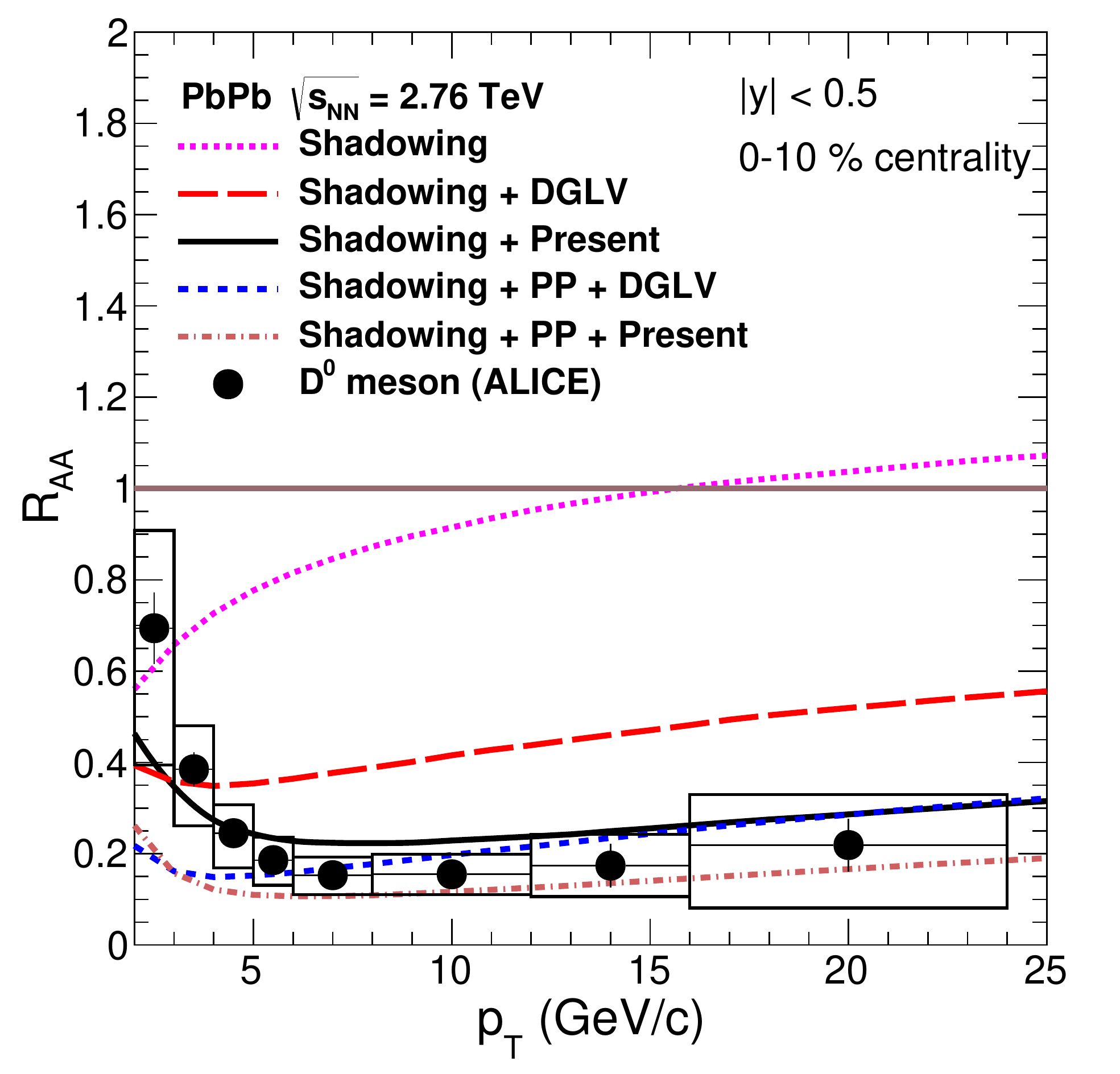}
\caption{(color online): Nuclear modification factor $R_{AA}$ of $D^{0}$ 
meson as a function of the transverse momentum $p_{T}$, obtained using energy loss 
(DGLV, Present, PP + DGLV and PP + Present calculation) and shadowing in PbPb collision 
at $\sqrt{s_{NN}}$= 2.76 TeV.
The calculations are compared with the CMS data of $D^{0}$ mesons \cite{Adam:2015sza}.}
\label{figure9_raa_0_10}
\end{figure}

\begin{figure}[htp]
\centering
\includegraphics[width=0.60\linewidth]{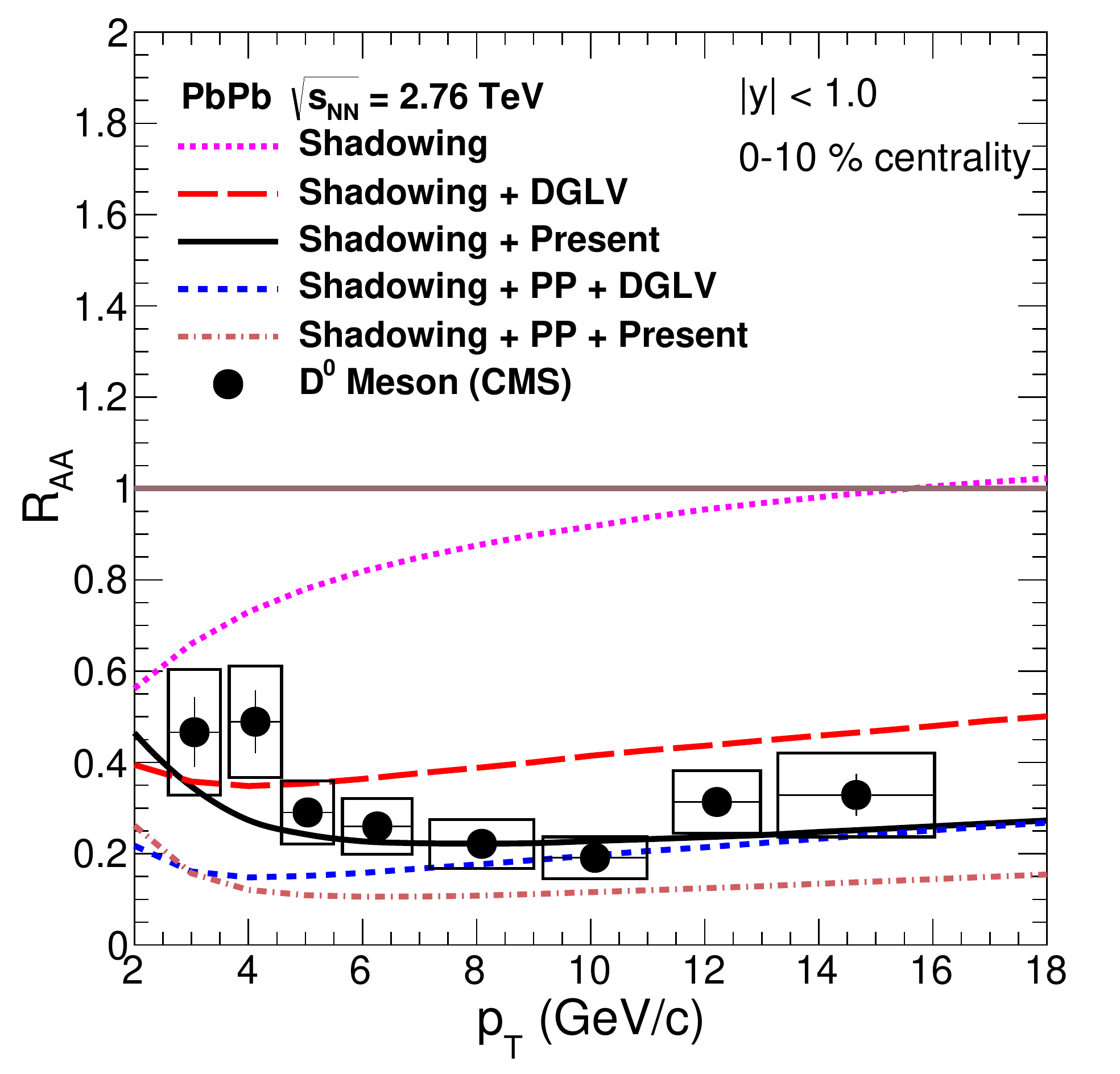}
\caption{(color online): Nuclear modification factor $R_{AA}$ of $D^{0}$ 
meson as a function of the transverse momentum $p_{T}$, obtained using energy loss 
(DGLV, Present, PP + DGLV and PP + Present calculation) and shadowing in PbPb collision 
at $\sqrt{s_{NN}}$= 2.76 TeV. The calculations are compared with the CMS data \cite{CMS:2015hca}.}
\label{figure10_raa_0_10}
\end{figure}

\begin{figure}[htp]
\centering
\includegraphics[width=0.60\linewidth]{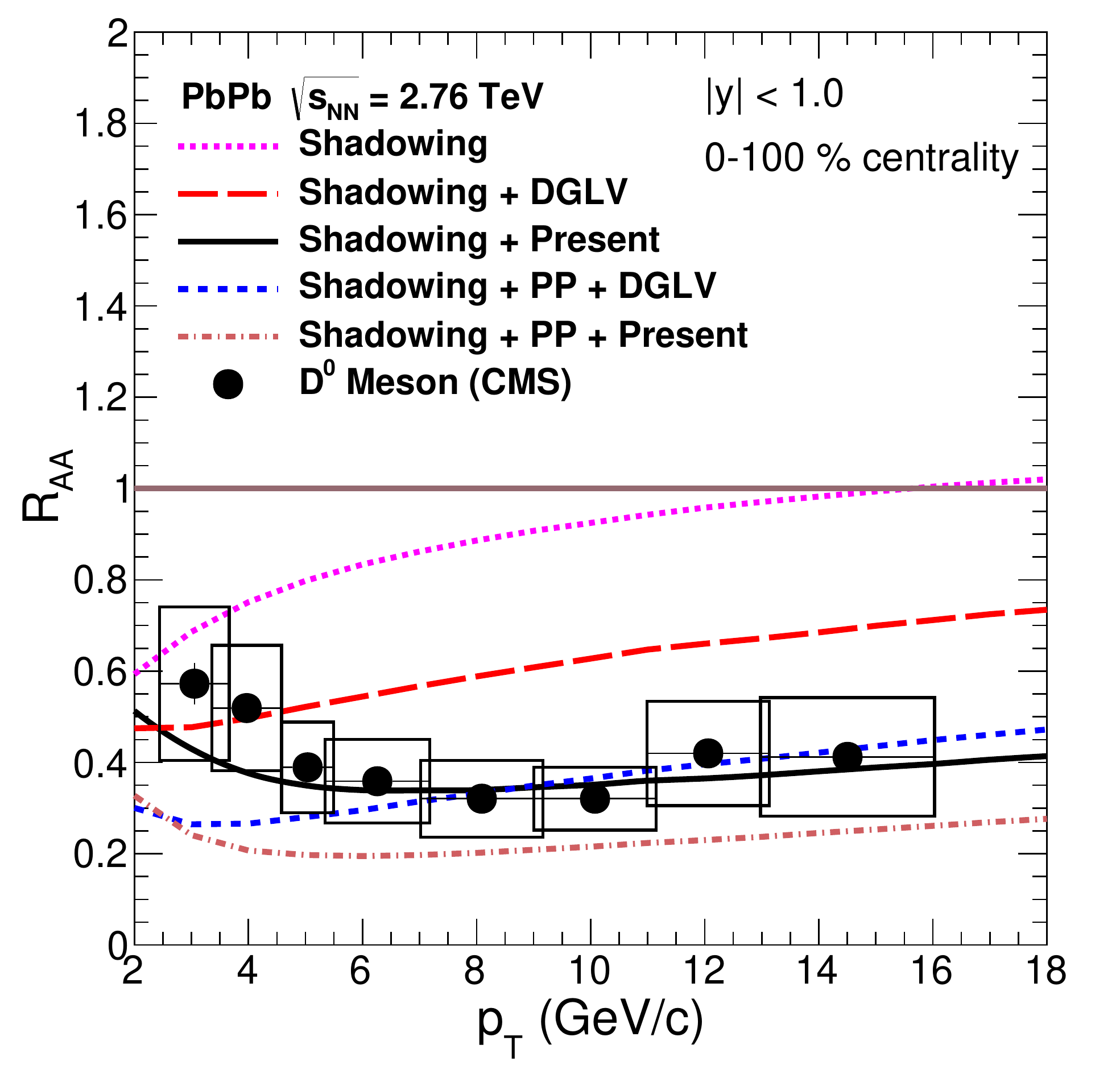}
\caption{(color online): Nuclear modification factor $R_{AA}$ of $D^{0}$ 
meson as a function of the transverse momentum $p_{T}$, obtained using energy loss 
(DGLV, Present,PP + DGLV and PP + Present calculation) and shadowing in PbPb collision 
at $\sqrt{s_{NN}}$= 2.76 TeV. The calculations are compared with the CMS data of $D^{0}$ mesons 
\cite{CMS:2015hca}.}
\label{figure11_raa_0_100}
\end{figure}

Figure \ref{figure9_raa_0_10} shows the nuclear modification factor $R_{AA}$ of 
$D^{0}$ as a function of the transverse momentum $p_{T}$, obtained by including shadowing 
and energy loss (DGLV, Present, PP + DGLV and PP + Present calculations) in the mid rapidity 
region $|y|<$ 0.5 for 0 - 10 $\%$ central PbPb collision at $\sqrt{s_{NN}}$= 2.76 TeV.
The calculations are compared with the ALICE data \cite{Adam:2015sza}.
 Figure \ref{figure10_raa_0_10} is the same as Fig.~\ref{figure9_raa_0_10} but for the 
case but for $|y|<$ 1.0, corresponding to CMS data \cite{CMS:2015hca}. 
Figure \ref{figure11_raa_0_100} is the same as Fig.~\ref{figure10_raa_0_10} but for 
the case in minimum bias PbPb collisions. 
The radiative energy loss by present formalism reproduces both the ALICE as well as
CMS data without adding collisional energy loss. The radiative energy loss by DGLV
added to the collisional energy loss by PP describes the data at high $p_{T}$. The 
sum of the radiative energy loss by present formalism and collisional energy loss by 
PP formalism overestimates the measured suppression of $D^{0}$ meson.

\section{Conclusion}

 In this work, first we calculate the $p_T$ spectra of heavy mesons in pp collision 
at $\sqrt{s}$ = 5.02 TeV using pQCD model and make a comparison with 
$D^{0}$ and $B^{+}$ meson measurements of CMS. The calculations reproduce the
shape of the $p_T$ spectra of the data reasonably well.
 A simple hydrodynamic picture is used for QGP evoluion during which the $p_T$
spectra of heavy quarks are modified due to collision and radiative energy loss.
The collisional energy loss is calculated using Peigne and Peshier formalism.
The radiative energy loss is obtained using two models namely reaction operator formalism 
and generalized dead cone approach. 
  The calculations are performed for the kinematic regions covered by ALICE and CMS 
measurements of $D$ meson $R_{AA}$ in PbPb collision at $\sqrt{s_{NN}}$ = 2.76 TeV and 
CMS measurements of $D^{0}$ and $B^{+}$  mesons $R_{AA}$ at $\sqrt{s_{NN}}$ = 5.02 TeV. 
  The radiative energy loss from generalized dead cone approach alone is sufficient 
to produce $D^{0}$ meson $R_{AA}$ at both the energies.
   For the case of $B^{+}$ meson, the radiative energy loss from generalized dead cone 
approach plus collisional energy loss gives good description of the data. It shows 
that collisional energy loss is siginificant for bottom quark.
  The radiative energy loss from DGLV formalism plus collisional 
energy loss gives good description of $D^{0}$ meson $R_{AA}$, but the 
sum of the radiative energy loss by DGLV formalism and collisional energy loss
underestimates the $B^{+}$ meson suppression.

\section*{Appendix}
\label{appendeix}

The Born cross section is given as  \cite{Beenakker:1988bq, Beenakker:1990maa}
\begin{eqnarray}
\sigma_{ij} = \frac{1}{64 \pi} K_{ij} \times ~ \sum |M_{ij}|^{2}~. 
\label{borncrosssectionone}
\end{eqnarray}
Here, $K$ is the color averaging factor. It is $1/(N^{2} - 1)^{2}$ for the gluon-gluon fusion process 
and is $1/N^{2}$ for the quark-antiquark annihilation process. The square of the amplitude averaged 
over the initial gluon polarization and color for the gluon gluon fusion is given as 
\cite{Beenakker:1988bq} 
\begin{eqnarray}
\sum |M_{gg}|^{2} &=& 2 ~ g^{4} \Big(C_{O} B_{O} + C_{K} B_{K} + C_{QED} B_{QED} \Big)~, \nonumber \\
C_{O} &=& N (N^{2} - 1)~,~ C_{K} = (N^{2} - 1) N^{-1}~~ \rm{and}~~ C_{QED} = 0~,  \nonumber \\
B_{QED} &=& \frac{t_{1}}{u_{1}} + \frac{u_{1}}{t_{1}} + 
\frac{4 m^{2} s}{t_{1} u_{1}} \Bigg( 1 - \frac{m^{2} s}{t_{1} u_{1}}\Bigg)~, \nonumber \\
B_{O} &=& \Bigg(1 - 2 \frac{t_{1} u_{1}}{s^{2}} \Bigg) B_{QED} ~ \rm{and}~ B_{K} = - B_{QED}~.
\label{matrixelementequationone} 
\end{eqnarray}
 The square of the amplitude averaged over the initial quark/antiquark spins and color for the 
quark-antiquark annihilation process is given as \cite{Beenakker:1990maa}
\begin{eqnarray}
\sum |M_{q\bar{q}}|^{2} &=& 4~ g^{4}~N~C_{F}~\Bigg(\frac{t^{2}_{1} + u^{2}_{1}}{s^{2}} + 
\frac{2 m^{2}}{s}\Bigg)~.
\end{eqnarray}
Here, $g(= \sqrt{4\pi\alpha})$ is the dimensionless coupling constant. 
$C_{F}\Big(= (N^{2} - 1)/(2 N)\Big)$ is the color factor corresponding to the fundamental 
representation of the quarks.

{\bf References}


\end{document}